\documentstyle[11pt,iopconf]{article}

\newcommand{\beq}{\begin{equation}}
\newcommand{\beqar}{\begin{eqnarray}}
\newcommand{\eeq}[1]{\label{#1}\end{equation}}
\newcommand{\eeqar}[1]{\label{#1} \end{eqnarray}}

\newcommand{\oript}{$\top$}
\newcommand{\oripp}{$^|_|$}
\newcommand{\oriet}{$\dashv$}
\newcommand{\oriep}{$||$}
\newcommand{\oriec}{
  \hspace{0.10ex}\makebox[0in][c]{$-$}\makebox[0in][c]{$|$}\hspace{0.90ex}}
\newcommand{\orip}{$\stackrel{\circ}{_|}$}
\newcommand{\orie}{$\circ \! |$}
\newcommand{\oris}{$\circ \! \bigcirc$}

\begin{document}
\renewcommand{\topfraction}{0.95}
\renewcommand{\textfraction}{0.05}
\pagestyle{myheadings}
\thispagestyle{empty}
\title{Collisions of Deformed
Nuclei and Superheavy-Element Production}
\author{
Akira Iwamoto\dag, Peter M\"{o}ller\dag\ddag\S\P,
J. Rayford Nix\P, and Hiroyuki Sagawa\ddag}
\affil{\dag Advanced Science Research Center,
Japan Atomic Energy Research Institute, Tokai,
Naka-gun, Ibaraki, 319-11 Japan}
\affil{\ddag Center for Mathematical Sciences, University of Aizu,
Aizu-Wakamatsu, Fukushima 965-80, Japan}
\affil{\S P. Moller Scientific Computing and Graphics, Inc.,
P. O. Box 1440, Los Alamos, NM 87544, USA}
\affil{\P Theoretical Division,
Los Alamos National Laboratory, Los Alamos, NM 87545, USA}
\beginabstract
A detailed understanding of complete fusion cross sections
in heavy-ion  collisions requires a consideration of the
effects of the deformation of the projectile and target.
Our aim here is to show that deformation
and orientation of the colliding nuclei have a very significant
effect on the fusion-barrier height and on the compactness of
the touching configuration.
To facilitate discussions of fusion configurations of deformed nuclei,
we develop a classification scheme
and introduce a notation convention for these configurations.
We discuss particular
deformations and orientations that lead to
compact touching configurations and to
fusion-barrier heights that correspond to fairly low excitation energies
of the compound systems. Such configurations should be the most
favorable for producing superheavy elements. We analyse a few
projectile-target combinations whose deformations allow favorable
entrance-channel configurations and whose proton and neutron numbers
lead to  compound systems in a  part of the superheavy region
where $\alpha$ half-lives are calculated to be
observable, that is, longer than 1 $\mu$s.

\endabstract

\section{Introduction}
The last five elements that   have been
discovered [1--5]
were all formed in cold-fusion reactions between spherical nuclei.
As the proton number increases, the cross section for heavy-element
production decreases. For example, element 107 was produced with a 167 pb
cross section [1],
whereas for element 111 the production cross section
was only 2--3 pb [5].
There is reason to suspect that few additional new elements can
be reached in reactions between spherical nuclei because of the strong
decreasing trend of the cross sections.

In fusion reactions where the number of protons in the
projectile and target add up to  about 100,  the overwhelming inelastic
cross-section component is fusion-fission. In a classical
picture a necessary condition for complete fusion and the formation
of a compound nucleus is that the fusing system evolves into
a configuration inside the fission saddle point in a multi-dimensional
deformation space [6--9].
In heavy-ion collisions  where the projectile and target are of roughly
equal size  and with a nucleon number $A$ above about 100,
the touching configuration is outside the fission saddle
point on the side of a steep
hill [10].
For energies
just above the Coulomb barrier this topographical feature results
in a trajectory that is deflected away from
the direction towards the spherical shape. Instead, it
leads from the touching configuration to the fission valley, so that
no compound-nucleus formation occurs.

There are two simple possibilities that immediately suggest themselves
to overcome the above limitation to compound-nucleus formation and
increase the cross section for heavy-element production. First, if
the projectile energy is increased, the trajectory will, for sufficiently high
energy, pass inside the fission saddle point. However, frictional
forces may make such trajectories difficult to realize. Second,
highly asymmetric touching configurations
may be sufficiently close to the ground-state shape of the compound
nucleus that the touching configuration is inside the
fission saddle point. Thus, these two simple principles would suggest
that to produce elements in the superheavy region one should select
highly asymmetric configurations and increase the projectile energy
above the Coulomb barrier. However, high excitation energies
and resultant high angular momenta
of the compound system may favor fission instead of de-excitation by
neutron emission. In the cold-fusion approach that led to the
identification of the five heaviest elements the very nature of
cold fusion leads to a low excitation energy of the compound system.
The entrance-channel configuration is also
fairly asymmetric and compact. However, the maximum cross section
for the production of the heaviest elements occurs at
sub-barrier energies as very rare, non-classical events.

Our discussion above revealed that from very general principles
one can expect that heavy-element production in heavy-ion reactions
is most favorable when the touching configuration is compact.
The excitation energy of the compound system should be high
enough to allow a trajectory inside the fission saddle point,
but as low as possible to reduce the fission branch of the compound
system. A spherical picture of nuclei in heavy-ion collisions allows
few new possibilities for very-heavy-element production beyond what
has already been accomplished. It is therefore of interest to
investigate if consideration of deformation will identify
entrance-channel configurations that have some possibility of
being  more favorable for heavy-element production than is expected
from the spherical picture.

To facilitate the discussion of deformed
fusion configurations we introduce a
classification scheme, notation and terminology.

\section{Fusion configurations of
  deformed nuclei:\protect\\
Classification, notation and terminology}

Obviously, the multi-dimensional fusion potential is a continuous
function of the incident direction and orientation of the
projectile nucleus and of the deformation of the
projectile and target. However, to allow the identification and discussion
of major physical effects it is useful to identify and
study a few limiting
situations.

\subsection{Limiting fusion configurations}

Our discussions of specific cases below will show
that for prolate shapes there are significant differences
in the fusion process depending on the sign of the hexadecapole
moment.
Nuclei with a
large negative hexadecapole moment
develop a neck which allows a close approach.
As a result the fusion configuration
for some orientations of the projectile-target combinations
is considerably more compact than the corresponding
configurations for shapes with large positive
hexadecapole moments.
Thus, we identify four limiting situations as far as deformations
are concerned. They are:
\begin{enumerate}
\item
Well-developed oblate shapes
\item
Spherical shape
\item
Well-developed prolate
shapes with large negative
hexadecapole moments $Q_4$
\item
Well-developed prolate
shapes with large  positive
hexadecapole moments $Q_4$
\end{enumerate}
Furthermore, we assume mass symmetry
and axial symmetry as this is consistent with the vast majority
of nuclear ground-state configurations.

In our studies here we use alternatively the
Nilsson perturbed-spheroid parameterization
$\epsilon$ [11] and the $\beta$  parameterization
 to  generate deformed nuclear shapes.
In the $\beta$ parameterization, assuming
axial symmetry, the radius vector $R(\theta,\phi)$
to the nuclear surface is defined by
\beq
R(\theta,\phi) = R_0\left[ 1 +\sum_{l=1}^\infty \beta_{l}Y_l^0(\theta,\phi)
\right]
\eeq{radvec}
where $R_0$ is deformation dependent so as to conserve the
volume inside the nuclear surface. The variation in $R_0$ due to
volume conservation is only a fraction of one percent.
The definition of the $\epsilon$ parameterization is more complicated.
A  recent, extensive presentation  is given in Ref.~[12].
One should note that large positive $Q_4$ corresponds to positive
$\beta_4$ but to negative $\epsilon_4$ and that
large negative $Q_4$ corresponds to negative
$\beta_4$ but to positive $\epsilon_4$.

As limiting orientations we consider only situations where
the projectile center is on the x, y or z axis of the target and
orientations of the projectile where the projectile symmetry axis is either
parallel to or perpendicular to the target symmetry axis.
Since we restrict ourselves to axial symmetry, configurations with
the projectile center  located on the x or y axis are identical.
If the projectile is located in the equatorial region of
the target it can be oriented in three major orientations,
and if it is located in the polar
region it can be oriented in two major orientations. Thus, for a particular
projectile-target deformation combination there are five
possible limiting configurations.

Because there are five orientations and three major types of deformations
for both projectile and target  there are 45 different
configurations when the projectile and target
are deformed and of unequal mass.
When the projectile and target are of equal
mass, one would at first sight expect 30 different configurations.
We later show that in the case of equal projectile and target mass
there are three pairs of configurations where the two configurations
in the pairs are identical. Therefore, there are in this case only 27 deformed
configurations that are different.
Situations where either the projectile or target
is deformed add another six configurations and, finally,
we designate a spherical target and a spherical projectile
as  a separate configuration. Thus, in our classification scheme
we find 34 configurations of projectile and target in
heavy-ion collisions that are different
also in the special case of equal projectile
and target mass.  For the case of unequal projectile and target mass
one may wish to count a total of 45 different deformed configurations,
for a total of 52 different fusion configurations.

We will in a separate study systematically
review the barrier parameters of
these  configurations for projectiles and targets throughout
the periodic system. Here, we will just discuss   a few
configurations with potential importance for very-heavy-element
production. However, to be able to simply and transparently refer
to any of the limiting configurations we start by introducing a notation
convention for deformed fusion configurations.

\subsection{Notation for deformed fusion configurations}

We denote a particular fusion configuration by
[P,T,O], where the three letters stand for
Projectile deformation, Target deformation, and relative
Orientation of the projectile-target combination.
For configurations where the projectile or target or both
are spherical, the number of different limiting orientations
is less than when both the projectile and target are deformed.
It is therefore most clear to introduce notation that distinguishes
between these possibilities.
The following values are
possible for the three entities P, T and O:
\begin{description}
\item[P and T]\mbox{ }\\
Oblate:\dotfill o\\
Spherical \dotfill s\\
Prolate with negative $Q_4$ \dotfill p$^-$\\
Prolate with positive $Q_4$ \dotfill p$^+$
\item[O] {\bf Spherical projectile and spherical target}\\
Spherical (s) \dotfill
    \oris

\item[O] {\bf Spherical-deformed projectile-target combination}\\
Polar (p) \dotfill
    \orip
\\
Equatorial (e) \dotfill
     \orie

\item[O] {\bf Deformed-deformed projectile-target configuration}\\
Polar-transverse (pt) \dotfill
    \oript \\
Polar-parallel (pp) \dotfill
     \oripp \\
Equatorial-transverse (et) \dotfill
            \oriet \\
Equatorial-parallel (ep) \dotfill
               \oriep   \\
Equatorial-cross (ec) \dotfill
                \oriec

\end{description}

We prefer the graphical short-hand notation given in the table above
for the different orientations, but we also provide in parenthesis
an alternative notation, based on letters only.

In Fig.~\ref{orisphe} we show the seven different configurations that can occur
with
\begin{figure}[t]
\begin{center}
\vspace{4.5in}
\caption[orisphe]{\baselineskip=12pt\small
The seven limiting touching configurations with
spherical projectiles.
The simplest configuration with a spherical target is in
the top row third from the left.
To the left of this configuration are configurations with prolate
target shapes
whereas  to the right are the two limiting configurations that occur
for oblate target shape.
The ratio between the projectile and target volume is 0.343.
The deformation is
$\beta_2=0.30$ and $\beta_4=0.11$
for p$^+$,
$\beta_2=0.24$ and $\beta_4 = - 0.09$ for p$^-$, and
$\beta_2 = - 0.25$ and $\beta_4=0.0$
for o shapes.
The arrows give the direction of the incident beam.
The nuclear symmetry axis is indicated by a thin line emerging from
the nuclear polar regions.}
\label{orisphe}
\end{center}
\end{figure}
a spherical projectile. We have sandwiched the familiar
spherical-projectile spherical-target case between the prolate-target and
oblate-target configurations in the top row so that the appearance
of the configurations evolves smoothly from the polar, spherical-prolate
positive-hexadecapole configuration [s,p$^-$,\orip]
on the extreme left to the polar, spherical-projectile oblate-target
configuration [s,o,\orip] on the far right.

In Fig. \ref{oripmpm} we show the five different limiting
orientations that occur for fixed target and projectile deformation
for the case where both target and projectile have prolate
deformation with large negative hexadecapole momnents.
In our classification scheme 45 different
\begin{figure}[t]
\begin{center}
\vspace{4.5in}
\caption[oripmpm]{\baselineskip=12pt\small
Five limiting touching configurations with
prolate, negative-hexadecapole projectiles
and targets.
Specifically
$\beta_2=0.24$ and $\beta_4=-0.09$.
The ratio between the projectile and target volume is 0.343.
Only the relative positions and orientations change between the
configurations. The arrows give the direction of the incident beam.
The nuclear symmetry axis is indicated by a thin line emerging from
the nuclear polar regions. }
\label{oripmpm}
\end{center}
\end{figure}
configurations occur when both the projectile and target
are deformed and of unequal mass.
In the case of equal projectile and target mass
the configuration [p$^+$,p$^-$,\oript] and [p$^-$,p$^+$,\oriet], for
example, are
identical.
Indeed, in this case all the configurations
[p$^-$,p$^+$,any] have a corresponding configuration
[p$^+$,p$^-$,any], and other similar correspondences also occur.
Therefore, for equal-mass projectile-target
combinations the configurations
[p$^+$,p$^+$,\oript],
[p$^-$,p$^-$,\oript] and
[o,o,\oript] are equivalent to
[p$^+$,p$^+$,\oriet],
[p$^-$,p$^-$,\oriet] and
[o,o,\oriet], respectively. This is the reason there are only
27 different configurations when the projectile and target are of
equal mass.

In Figs. \ref{orisphe} and \ref{oripmpm} we use the $\beta$ parameterization
to describe the nuclear shape. Volume conservation has not
been applied in these and subsequent figures of nuclear shapes,
but this is an
insignificant approximation since volume conservation only changes
$R_0$ by fractions of a percent for the deformations considered. However, in
energy
calculations it is essential to include volume conservation, as we do
in our calculations here.
As representative deformations we make the following choices.
As the prolate--positive hexadecapole deformation
p$^+$ we choose $\beta_2=0.30$ and $\beta_4=0.11$.
This corresponds to the experimentally determined deformation of
$^{154}$Sm [12]. The prolate-negative hexadecapole
deformation
p$^-$ is chosen as $\beta_2=0.24$ and $\beta_4=-0.09$, corresponding
to the experimentally determined deformation of $^{186}$W [12].
Finally, as a representative oblate deformation we have selected
$\beta_2=-0.25$ and $\beta_4=0.0$. The ratio between $R_0$ of the
projectile and target  is 0.7.

\section{Deformation and heavy-ion collisions}

Although the implications of deformation on cross sections for
superheavy-element production have not been very extensively
considered so far, deformation certainly is already known to affect
fusion cross sections leading to somewhat lighter compound systems.
For example, a clear signature of the importance of deformation
effects in heavy-ion reactions is the enhancement of sub-barrier
fusion cross sections, for which deformation often plays a major role.
It may be useful to observe that the designation sub-barrier is
somewhat of a misnomer.  An implicit assumption behind this
designation is that both projectile and target nuclei are spherical.
Furthermore, if the measured cross section at energies below the
maximum of this assumed spherical fusion barrier is higher than the calculated
cross section for this configuration then the term {\it enhanced
sub-barrier fusion} is used.  In a more realistic picture one can in
many cases show that (1) the energy is not sub-barrier and (2) the
measured cross section is not enhanced.  To illustrate these features
we select the reaction $^{16}{\rm O}+\mbox{}^{154}$Sm.

\subsection{Deformation and the fusion potential-energy surface}

We present in Table 1 four fusion-barrier quantities for particular
orientations between the projectile and target. Each line corresponds to
\begin{table}[t]
{\small
\begin{center}
\caption[taberr]{\baselineskip=12pt\small
Comparison of entrance-channel fusion configurations.
When the sign $<$ is given in the column for $R_{\rm max}$
and $>$ is given in the column for $V_{\rm max}$
it means that the maximum of the fusion barrier occurs inside
the touching point and consequently is higher than
the potential of the touching configuration.
\\}
\begin{tabular}{rrrrrrrrrcrrrrr}
\hline\\[-0.07in]
\multicolumn{4}{c}{Target}      & &
\multicolumn{5}{c}{Projectile} & &
\multicolumn{4}{c}{Barrier}
\\[0.08in]
\cline{1-4} \cline{6-10} \cline{12-15}\\[-0.07in]
             &
\multicolumn{1}{c}{$\epsilon_2$} &
\multicolumn{1}{c}{$\epsilon_4$} &
\multicolumn{1}{c}{$\epsilon_6$} &
             &
             &
\multicolumn{1}{c}{$\epsilon_2$} &
\multicolumn{1}{c}{$\epsilon_4$} &
\multicolumn{1}{c}{$\epsilon_6$} &
\multicolumn{1}{c}{Or.}&
             &
\multicolumn{1}{c}{$R_{\rm max}$}&
\multicolumn{1}{c}{$V_{\rm max}$}&
\multicolumn{1}{c}{$R_{\rm t}$}&
\multicolumn{1}{c}{$V_{\rm t}$}
\\
  &
  &
  &
  &
  &
  &
  &
  &
  &
\multicolumn{1}{c}{ }  &
  &
\multicolumn{1}{c}{(fm)}  &
\multicolumn{1}{c}{(MeV)} &
\multicolumn{1}{c}{(fm)}  &
\multicolumn{1}{c}{(MeV)}
\\[0.08in]
\hline\\[-0.07in]
 $^{154}$Sm &
0.000       &
0.000       &
0.000       &
           &
 $^{16}$O &
0.000       &
0.000       &
0.000       &
\oris       &
           &
$10.54$       &
$62.21$       &
9.14     &
56.22       \\
%
%
 $^{154}$Sm &
0.250       &
$-0.067$       &
0.030       &
           &
 $^{16}$O &
0.000       &
0.000       &
0.000       &
\orie       &
           &
$10.10$       &
$63.29$       &
8.80     &
57.90       \\
%
%
 $^{154}$Sm &
0.250       &
$-0.067$       &
0.030       &
           &
 $^{16}$O &
0.000       &
0.000       &
0.000       &
{\footnotesize \orip}       &
           &
$11.87$       &
$57.18$       &
10.67     &
53.34       \\
%
%
 $^{150}$Nd &
0.000       &
0.000       &
0.000       &
           &
 $^{150}$Nd &
0.000       &
0.000       &
0.000       &
\oris       &
           &
$<$       &
$>$       &
12.33      &
379.10       \\
%
%
 $^{150}$Nd &
0.225       &
$-0.067$       &
0.025       &
           &
 $^{150}$Nd &
0.225       &
$-0.067$       &
0.025       &
{\footnotesize
\oriec }       &
           &
$<$          &
$>$          &
11.74      &
390.96
\\[1ex]
%
%
 $^{150}$Nd &
0.225          &
$0.200$        &
$-0.100$       &
               &
 $^{150}$Nd    &
0.225          &
$0.200$        &
$-0.100$       &
{
\oriec }  &
           &
11.69      &
399.51     &
10.29      &
383.98
\\
%
%
 $^{150}$Nd &
0.225          &
$0.100$        &
$-0.044$       &
               &
 $^{150}$Nd    &
0.225          &
$0.100$        &
$-0.044$       &
{
\oriec  }   &
           &
11.66      &
396.73     &
10.86      &
392.38
\\
%
%
 $^{186}$W &
0.208          &
$0.100$        &
$-0.044$       &
               &
 $^{110}$Pd    &
0.200          &
$0.027$        &
$-0.013$       &
{
\oript }       &
           &
$<$      &
$>$     &
12.29     &
358.13
\\
%
%
 $^{186}$W &
0.208          &
$0.100$        &
$-0.044$       &
               &
 $^{110}$Pd    &
0.200          &
$0.027$        &
$-0.013$       &
{\footnotesize
\oripp }       &
           &
$<$     &
$>$     &
13.46     &
342.61
\\
%
%
 $^{186}$W &
0.208          &
$0.100$        &
$-0.044$       &
               &
 $^{110}$Pd    &
0.200          &
$0.027$        &
$-0.013$       &
{
\oriet }       &
           &
$<$      &
$>$     &
12.15     &
359.01
\\[1ex]
%
%
 $^{186}$W &
0.208          &
$0.100$        &
$-0.044$       &
               &
 $^{110}$Pd    &
0.200          &
$0.027$        &
$-0.013$       &
{\footnotesize
\oriep }       &
           &
11.69      &
375.12     &
10.99     &
372.84
\\
%
%
 $^{186}$W &
0.208          &
$0.100$        &
$-0.044$       &
               &
 $^{110}$Pd    &
0.200          &
$0.027$        &
$-0.013$       &
{\footnotesize
\oriec }       &
           &
11.69      &
376.20     &
10.99     &
374.14
\\
%
%
 $^{186}$W &
0.000          &
$0.000$        &
$0.000$       &
               &
 $^{110}$Pd    &
0.000          &
$0.000$        &
$0.000$       &
\oris      &
           &
$<$      &
$>$     &
12.18     &
361.10
\\
%
%
 $^{192}$Os&
0.142          &
$0.073$        &
$-0.032$       &
               &
 $^{104}$Ru    &
0.233          &
$-0.013$        &
$0.012$       &
\oriec      &
           &
$11.72$      &
$367.82$     &
11.22     &
367.11
\\[0.08in]
\hline
\end{tabular}\\[3ex]
\end{center}
}
\end{table}
one orientation and one incident direction.
The first eight columns specify the
projectile and target nuclei and the deformation used
for these nuclei in the calculation of
the fusion barrier.
The shapes of the projectile and target are given in the Nilsson
perturbed-spheroid parameterization [11].
The next column gives the relative orientation of
projectile and target
in the notation introduced above.
The last four
columns indicate (1) the distance between the centers of the
projectile and target
at the maximum of the barrier, (2) the maximum of the fusion
barrier, (3) the center-of-mass distance when the projectile and target
just touch and (4) the fusion-barrier height at this point.

The first three lines of Table 1 show fusion-barrier data for the
reaction $^{16}{\rm O}+\mbox{}^{154}$Sm.  In the first line of the table we
show, for reference, the calculated barrier parameters for a
hypothetical spherical target shape. The second line gives the
fusion-barrier parameters for the configuration [s,p$^+$,\orie]
corresponding to the equatorial plane $z=0$
and the third line
corresponds to the potential in the  [s,p$^+$,\orip] configuration.

\subsection{Deformation and fusion cross sections}

In Fig.~\ref{crsmo} the measured and
\begin{figure}[t]
\begin{center}
\vspace{3.5in}
\caption[crsmo]{\baselineskip=12pt\small
Calculated fusion cross sections for the reaction
$^{16}{\rm O} + ^{154}$Sm,
compared to experimental data.
The solid curve corresponds to the calculated fusion cross section obtained
when the shape of the target corresponds
to the calculated ground-state shape. The long-dashed curve is the cross
section obtained for a hypothetical spherical  target.
The arrows show the fusion-barrier height in the polar direction
(p), the equatorial plane (e), and the barrier height for a
hypothetical
spherical target (s).
Both the curves and the arrows have
been translated in energy by $E_{\rm tran} = - 3.1 $ MeV from
their calculated values.
}
\label{crsmo}
\end{center}
\end{figure}
calculated cross sections corresponding to the reaction
$^{16}{\rm O} + ^{154}$Sm are presented.
The deformed fusion potential is obtained in a
model calculation with no free parameters
and is the sum of the nuclear and Coulomb potentials according to Ref.
[12] and
a centrifugal barrier term, which is treated
in the spherical limit.
The calculated cross section is from a study~[13]
of fusion cross sections in reactions
of spherical projectiles and deformed targets.  It has no free
parameters except a simple translation in energy of the calculated
cross-section curves.
The cross section is obtained by integrating over angle the
transmission
coefficients which are obtained by calculating the barrier
penetrability at each angular momentum by use of the WKB
approximation.
The deformation parameters of
the target are obtained from a mass calculation [14].
Obviously there are large deformation effects both in the potential
energy and in the fusion cross section. Clearly our model,
incorporating significant aspects of
deformation, accounts well for the ``enhancement'' of the
cross section relative to the fusion cross section
obtained for a hypothetical spherical target,
at least for energies down to the Coulomb barrier in
the polar direction.

\subsection{Gentle fusion?}

Because the evaporation residue cross sections
in cold fusion between spherical projectiles and targets
drop so strongly
towards heavier nuclei, N{\"{o}}renberg [15,16]
suggested that ``gentle fusion''
of two well-deformed rare-earth nuclei in an equatorial-cross
orientation \oriec\
should be investigated because, he stated,
``this orientation leads to
the most compact touching configuration
out of all possible orientations of the two deformed nuclei.'' Consequently,
the
evaporation-residue cross sections may be sufficiently large to allow
detection.

We first observe that according to our calculations [14],
only the lightest nuclei in the rare-earth region would lead to
compound systems with $\alpha$ half-lives over 1 ${\mu}$s, which is
the approximate transit time from the target to detection area in the
SHIP experimental setup. Already the reaction $^{160}{\rm
  Gd}+\mbox{}^{160}$Gd$\rightarrow \mbox{}^{320 - {\rm xn}}128 + {\rm
  xn}$ leads to nuclei where the calculated
[14,17] $\alpha$-decay half-lives are less than
about 0.01 $\mu$s. To study the concept of gentle fusion we must
therefore select a reaction in the beginning of the rare-earth region,
so we choose the reaction $^{150}{\rm Nd}+\mbox{}^{150}$Nd to illustrate
N\"{o}renberg's suggestion.  We show the configuration of two
$^{150}$Nd nuclei  with calculated ground-state
shapes in Fig.~\ref{ndndgent}. The configuration is
[p$^+$,p$^+$,\oriec]
\begin{figure}[t]
\begin{center}
\vspace{3.5in}
\caption[ndndgent]{\baselineskip=12pt\small
Touching configuration of $^{150}{\rm Nd}+\mbox{}^{150}$Nd  with the nuclear
shapes taken to be the calculated [14] ground-state
shape;
that is,  the configuration is $[{\rm p}^+,{\rm p}^+,$\oriec $]$.
The arrow gives the direction of the incident beam.
Fusion-barrier parameters for this configuration/direction
are given on line 5 of Table 1.
}
\label{ndndgent}
\end{center}
\end{figure}
and is the one proposed by N{\"{o}}renberg as favorable
for SHE production. Calculated fusion-barrier data for the
hypothetical spherical case and the configuration in
Fig.~\ref{ndndgent} are found in Table 1, on lines 4 and 5,
respectively.

It is clear
that the fusion
configuration \oriec\
suggested by N{\"{o}}renberg is limited to
[p$^+$,p$^+$,\oriec] configurations, since
projectiles and targets must be chosen from the beginning of the
rare-earth region.
This configuration is not particularly compact relative to
a collision between similar-size spherical nuclei,
as is clear from  Fig.~\ref{ndndgent} and  Table 1.
Indeed, because of the large negative $\epsilon_4$ of the ground
state, which results in a bulging equatorial region
and a large positive hexadecapole moment, the configuration
in Fig.~\ref{ndndgent} is quite similar to the spherical configuration.
 This observation is supported by the
quantitative results in Table 1: the distance between mass centers of
the gentle fusion configuration is 11.74 fm, only 0.59 fm more compact
than the spherical configuration.

The idea that  configurations where deformed nuclei
touch each other in the equatorial regions are more compact than
some other configurations  and may therefore be
favorable for SHE production is not new. It was for instance mentioned
in Ref.~[12] in a discussion of the reaction
$^{48}{\rm Ca}+\mbox{}^{248}$Cm, and we will return to this reaction below.
Clearly, the fusion barrier for deformed systems
along a one-dimensional path will be very different
in the polar direction and in an
equatorial direction.
When the projectile is deformed the fusion barrier
will also depend strongly on the
orientation of the incident deformed projectile.

It is obvious that when colliding heavy ions have well-developed
prolate deformation, then the most compact configurations occur
when the point of touching is in the equatorial region of both
nuclei. Which relative orientation  of the two
nuclei, \oriec\ or {\oriep}, is the most
favorable is perhaps not known at present. However, the orientation
suggested by N\"{o}renberg is one possible favorable configuration,
but its properties will depend strongly on the value of the
hexadecapole deformation, that is, in our case on the value of the
deformation parameter $\epsilon_4$. Large negative values of
$\epsilon_4$ correspond to bulging equatorial regions, whereas
positive values lead to neck formation. We now look at the latter, more
compact configurations.

\subsection{Hugging fusion!}

To clearly illustrate the effect of large positive values of
the deformation parameter $\epsilon_4$ we first study an example where we for
clarity exaggerate somewhat the effect.
We show in Fig.~\ref{ndndhugp} the configuration in
Fig.~\ref{ndndgent}, with one change, namely we select $\epsilon_4$ and
\begin{figure}[t]
\begin{center}
\vspace{3.5in}
\caption[ndndhugp]{\baselineskip=12pt\small
Touching configuration of $^{150}{\rm Nd}+\mbox{}^{150}$Nd for
hypothetical nuclear shapes with a large positive $\epsilon_4$
and a choice of $\epsilon_6$ that further develops the waistline;
that is,  the configuration is $[{\rm p}^-,{\rm p}^-,$\oriec $]$.
The arrow gives the direction of the incident beam.
Fusion-barrier parameters for this configuration/direction
are given on line 6 of Table 1.
}
\label{ndndhugp}
\end{center}
\end{figure}
$\epsilon_6$ so that a well-developed neck results.
The configuration is [p$^-$,p$^-$,\oriec]. The corresponding
calculated fusion-barrier parameters are listed on line 6 of Table 1.
This hypothetical shape is presented to show the effect
of a well-developed neck on the fusion barrier
and touching configuration. Clearly this configuration is
very different from both the spherical configuration and
the gentle configuration and quite compact. Similar
configurations with necks in the equatorial regions
instead of bulging midsections could favor a large
cross section for complete fusion. Because the nuclei
``grab'' each other we call this configuration corresponding to
this specific orientation and where both
projectile and target exhibit some neck formation {\it
hugging fusion}. In our classification scheme hugging
fusion corresponds to
the [p$^-$,p$^-$,\oriec]
class of touching fusion configurations.
The $\epsilon_4$  deformation value
selected to clearly show this principle is probably unrealistically
large. However, large positive $\epsilon_4$ deformations occur in the
end of the rare-earth region. To compare the effect of a realistic
positive value of $\epsilon_4$  with the effect of a large negative
$\epsilon_4$ we apply the deformation calculated [14] for
$^{186}$W to $^{150}$Nd and obtain the
fusion barrier given on line 7 of Table 1.
We see that the distance between mass centers of this configuration is
only 10.86 fm, that is, 1.47 fm more compact than the spherical
configuration and 0.88 fm more compact than a configuration with
a large negative $\epsilon_4$. To exploit the enhancement of
the evaporation-residue cross section that we expect in the hugging
configuration [p$^-$,p$^-$,\oriec]
 we must find
suitable projectiles and targets   with large positive $\epsilon_4$
ground-state deformations that lead to superheavy elements with
half-lives that are sufficiently long that the evaporation residues
are observable.

\section{Heavy-ion reactions for distant superheavy-element
  production}

The most stable nuclei on the superheavy island are predicted to occur
in the vicinity of $^{288\mbox{\rm --}294}$110
even though the magic proton number
in this region is calculated to be 114 [17].
However, nuclei at some considerable distance away from
the center of the island are calculated to be sufficiently long-lived
to allow observation after formation; that is, they are predicted to
have half-lives in excess of 1~$\mu$s. We refer to  elements with
proton number larger than 114 as distant superheavy elements.
We now look at some heavy-ion reactions that may lead to
this far part of the superheavy island.

\subsection{Hugging fusion candidates for distant superheavy-element
production}

Above we noted that to achieve very compact configurations of deformed
nuclei one should find projectiles and targets  with large positive
values of the $\epsilon_4$ deformation parameter. Clearly then, the
best candidates for a stable target
above proton number 50 would  be nuclei near the
end of the rare-earth region. To be specific, we select $^{186}$W as a
target in our first example. For this nucleus, calculations [14]
give $\epsilon_4=0.100$ and
$\epsilon_6=-0.044$. The large negative value
of $\epsilon_6$ also
contributes to the development of a neck. A suitable projectile that
would take us to the region of distant superheavy elements would then
be $^{110}$Pd leading to the compound system $^{296}$120. The hugging
configuration for this choice is shown from four different angles in
Fig.~\ref{wpd4}. The fusion barrier for the hugging configuration
\begin{figure}[t]
\begin{center}
\vspace{3.5in}
\caption[wpd4]{\baselineskip=12pt\small
Touching configuration of $^{110}{\rm Pd}+\mbox{}^{186}$W for
calculated ground-state shapes viewed from four different angles.
The shapes used are the calculated ground-states shapes, so the
configuration is $[{\rm p}^-,{\rm p}^-,$\oriec $]$.
The arrows and $\bigotimes$ sign give the direction of the incident beam.
Fusion-barrier parameters for this configuration/direction
are given on line 12 of Table 1.
}
\label{wpd4}
\end{center}
\end{figure}
[p$^-$,p$^-$,\oriec] is  listed  on line 12 of Table 1,
where we to illustrate the orientation effect on the fusion
barrier also  list the barrier parameters for the four other deformed
configurations
[p$^-$,p$^-$,\oript],
[p$^-$,p$^-$,\oripp],
[p$^-$,p$^-$,\oriet] and
[p$^-$,p$^-$,\oriep] on lines 8--11.
These five deformed configurations also appear in Fig.~\ref{oripmpm}
for slightly different projectile-target sizes and deformations.
The table listing on lines 8--12 is in the order the configurations
occur in
Fig.~\ref{oripmpm}.
In Table 1 we also list on line 13, for reference,
the  barrier parameters for the [s,s,\oris] configuration.

To make an estimate of the decay properties of the compound system we make
the following assumptions.
The heavy-ion reaction takes place at the fusion-barrier energy.
We do not calculate the branching ratio between fusion-fission
and complete fusion, but are primarily interested in studying
the alpha-decay rates of the compound nuclei that possibly do not
fission but de-excite by neutron emission.
One expects of course that at high excitation energy some washing
out of shell effects has taken place and that
$\Gamma_{\rm f}/\Gamma_{\rm n}$ is large. It is a remaining,
important problem to calculate this quantity.
We assume that
neutrons are emitted as long as energetically possible.
The $Q$-values and masses required for these calculations are obtained
from Ref.~[14]. The $\alpha$-decay half-lives are
calculated as discussed in Ref.~[17].
With these assumptions we find for the reaction
and configuration [p$^-$,p$^-$,\oriec]
shown in Fig.~\ref{wpd4} at a center-of-mass energy
equal to  the Coulomb barrier energy listed on line 12 in Table
1 that two neutrons are emitted. Thus
\beq
^{110}{\rm Pd}+ \mbox{}^{186}{\rm W}
\rightarrow \mbox{}^{296}120^*
\rightarrow \mbox{}^{294}120 + {\rm 2n}
\eeq{react1a}
where the compound nucleus has an excitation energy of 35.04 MeV
before neutron emission. The $\alpha$-decay-chain half-lives and
$Q$-values are shown in Fig.~\ref{cpdw}. Although the first few
\begin{figure}[t]
\begin{center}
\vspace{4.5in}
\caption[cpdw]{\baselineskip=12pt\small
Calculated $Q$-values for $\alpha$ decay and corresponding calculated
half-lives for the decay chain starting
at $^{294}120$.   }
\label{cpdw}
\end{center}
\end{figure}
decays are calculated to be only a few $\mu$s, these decays should
be within the detection limit of SHIP. Fission half-life calculations
are characterized by large uncertainties [17], but
the calculated ground-state microscopic corrections in the
region of the compound system are about $-7$~MeV, so one expects a
fission barrier about this high in this region of nuclei.
Such a high barrier would probably be associated with fission
half-lives that are longer than the calculated $\alpha$ half-lives
down to about element 104 for all the decay chains considered here.

We have also considered the reaction
\beq
^{104}{\rm Ru}+ \mbox{}^{192}{\rm Os}
\rightarrow \mbox{}^{296}120^*
\rightarrow \mbox{}^{294}120 + {\rm 2n}
\eeq{react1b}
The barrier parameters are listed  on line 14 in Table 1. A
beam energy equal to the Coulomb barrier value of 367.82 MeV leads to
a compound-nucleus excitation energy of 34.06 MeV, which is about 1 MeV lower
than in the reaction~(\ref{react1a}), and consequently to the
same $\alpha$-decay sequence after 2n emission.

\section{Summary}

In heavy-ion collisions between deformed projectiles and targets
we have shown that the fusion reaction
depends strongly on the relative orientation of the
projectile and target.
Both the fusion-barrier height and the compactness of the
touching configuration are  so strongly affected that
a variation of relative orientation may have a similar impact as
varying the projectile  and/or target nuclear species. Therefore,
a detailed consideration of deformation is necessary in both theory
and experimental work so that we can understand more about
the many features of heavy-ion reactions between deformed nuclei.
To facilitate such studies we have introduced a classification scheme
of deformed fusion configurations.

Systematic experimental work on understanding cold-fusion reactions
and associated cross sections for evaporation-residue formation
and parallel investigations of microscopic nuclear-structure models
have over the last 20 years or so led to the discovery of five new
elements on the side of the superheavy island closest to us.  Similar
or more extensive work will be required to describe in detail the
fusion reactions between two deformed nuclei. However, the reward may
be access to the far side of the superheavy island.  Of particular
interest is to study how the high charge numbers of these nuclei
affect nuclear and atomic properties. Above we have given a few
examples of heavy-ion reactions that could serve as particularly
suitable starting points for exploring both theoretically and
experimentally the new physics of deformed heavy-ion reactions, and
possibly the new physics of the far side of the superheavy island.
In particular we have suggested that a few special fusion configurations may
be especially favorable for forming superheavy elements. In hot fusion,
we suggest as most favorable an asymmetric projectile-target
combination in the {\it hugging} configuration
$[{\rm p}^-,{\rm p}^-,$\oriec $]$.

A more extensive discussion of the ideas presented here may be found
in Ref.~[18].

This work was supported by the Japan Atomic Energy Research
Institute
and by the U.~S.\ Department of
Energy.


\begin{small}

\end{small}
\end{document}